\documentclass[%
aip,
amsmath,amssymb,
reprint,%
]{revtex4-1}

\usepackage{dcolumn}
\usepackage{bm}

\usepackage[utf8]{inputenc}
\usepackage[T1]{fontenc}
\usepackage{mathptmx}

\usepackage{graphicx}
\usepackage{caption}
\usepackage{subcaption}

\usepackage{xcolor}
\usepackage{xfrac}
\usepackage{float}

\usepackage{comment}

\usepackage[mathlines]{lineno}

\usepackage{etoolbox}
\makeatletter
\patchcmd\linenumberpar{\@LN@parpgbrk}{\penalty\@LN@parpgpen\relax}{}{}
\makeatother

\begin{document}

\title{Capillary driven flows in microgravity}

\author{Domenico Fiorini}
\affiliation{%
	von Karman Institute for Fluid Dynamics, Waterloosesteenweg 72, Sint-Genesius-Rode, Belgium
}%
\affiliation{%
	KU Leuven, Dept. of Materials Engineering, Leuven 3001, Belgium
}%
\author{Alessia Simonini}
\affiliation{%
	von Karman Institute for Fluid Dynamics, Waterloosesteenweg 72, Sint-Genesius-Rode, Belgium
}%
\author{Johan Steelant}
\affiliation{%
	ESTEC-ESA, Keplerlaan 1, Noordwijk, The Netherlands
}%
\author{David Seveno}
	\affiliation{%
	KU Leuven, Dept. of Materials Engineering, Leuven 3001, Belgium
}%
\author{Miguel Alfonso Mendez}
\email{mendez@vki.ac.be}
\affiliation{%
	von Karman Institute for Fluid Dynamics, Waterloosesteenweg 72, Sint-Genesius-Rode, Belgium
}%

\date{\today}
	
\begin{abstract}
This work investigates the capillary rise dynamics of highly wetting liquids in a divergent U-tube in the microgravity conditions provided by 78th European Space Agency (ESA) parabolic flight. This configuration produces a capillary-driven channel flow. We use image recording in backlight illumination to characterize the interface dynamics and dynamic contact angle of HFE7200 and Di-Propylene Glycol (DPG). For the case of HF7200, we complement the interface measurements with Particle Tracking Velocimetry (PTV) to characterize the velocity fields underneath the moving meniscus. In the experiments with DPG, the liquid column reaches different heights within various experiments, and the measurements show a sharp reduction of the meniscus curvature when the contact line moves from a pre-wet to a dry substrate. In experiments with HFE7200, the interface always moves on a pre-wet surface. Yet a curvature reduction is observed due to the inertial forces on the underlying accelerating flow. The PTV measurements show that the distance from the interface within which the velocity profile adapts to the meniscus velocity shortens as the interface acceleration increases.


\end{abstract}
	
\maketitle


\section{Introduction}\label{sec:intro}
The capillary behaviour of a liquid results from the shape of the gas-liquid interface and the wetting interaction with the substrate. The dynamic contact angle between the interface and the substrate models this interaction and sets the boundary condition for any model of the interface shape.

Early investigations by \citet{voinov1976hydrodynamics} and \citet{kistler1993hydrodynamics} showed that the dynamic contact angle depends on the hydrodynamics of the flow near the contact line and the capillary number: Ca $=\mu U_{CL} \sigma^{-1}$ (with $\mu$, $\sigma$ the liquid's dynamic viscosity and gas-liquid surface tension and $U_{CL}$ the velocity of the interface's contact line with the substrate). These studies are primarily based on the assumption of a steady, two-dimensional, viscous-dominated flow near the contact line.
Recent work seeks to extend the description of capillary-driven flows to more general scenarios where gas-liquid interfaces are dominated by inertial forces \cite{varma2021inertial,fiorini2022effect,fiorini2023PRF}. These investigations show the impact of the flow further away from the contact line on the interface shape and its motion. In the absence of gravity, inertial forces can also affect the flow motion initiated by capillarity, particularly when viscous forces are negligible. To study this possibility, \citet{levine2015surface} analyze capillarity and dynamic contact angle in microgravity environments, where the motion of the gas-liquid interface primarily results from the balance of capillary and inertial forces. To this end, the capillary dynamics of fluids for micro-gravity applications are investigated using prototype test cases such as the capillary rise in tubes and micro-gravity platforms. Among these studies, the pioneering investigations by \citet{stange2003capillary} and \citet{gerstmann2007dynamic} characterized the capillary rise of various highly wetting liquids in a drop tower. These studies compare the capillary rise with a theoretical model based on the dynamic contact angle correlation provided by \citet{jiang1979correlation} and demonstrate that the interface is spherical under microgravity conditions. However, no direct measurement of the contact angle was reported, and the characterization of the gas-liquid interface was limited to qualitative evaluation of video recordings.

In this work, we extend the previous experimental studies on capillary flow under microgravity conditions with three main contributions. First, we present detailed measurements of the gas-liquid interface evolution, allowing to retrieve the evolution of the dynamic contact angle and thus analyze its relation to the contact-line velocity and acceleration. Second, we complement the interface characterization with image velocimetry underneath the interface to analyze the interaction between velocity fields and interface shape. Finally, we conducted our experiments in a novel configuration, which we refer to as the Divergent U-tube (DUT) and a parabolic flight platform. Compared to the experiments in \citet{stange2003capillary}, this configuration offers a longer microgravity phase while at the same time allowing much larger accelerations than those observed in capillary experiments on ground. Moreover, it offers the advantages of a U-tube configuration \citep{fricke2023analytical}, namely avoiding the challenges in modelling the entrance effect and allowing for deriving much simpler reduced order models \citep{fiorini2023eucass}. This work extends the preliminary results in \citet{fiorini2023MST} by adding new experiments, a second liquid, and a more advanced characterization of the velocity fields.



\section{The DUT Experiment: concept and scales}\label{sec:concept}

\begin{figure}[ht!]
       \subfloat[\label{subfig:interface1g}]{\includegraphics[width=0.25\textwidth, trim = 0.5cm 0cm 0.5cm 0cm, clip]{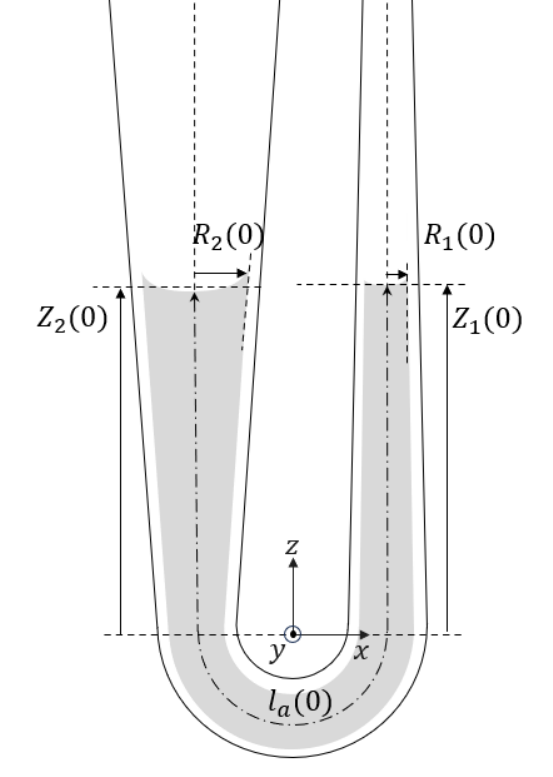}}
  \subfloat[\label{subfig:interface0g}]{\includegraphics[width=0.25\textwidth, trim = 0.5cm 0cm 0.5cm 0cm , clip]{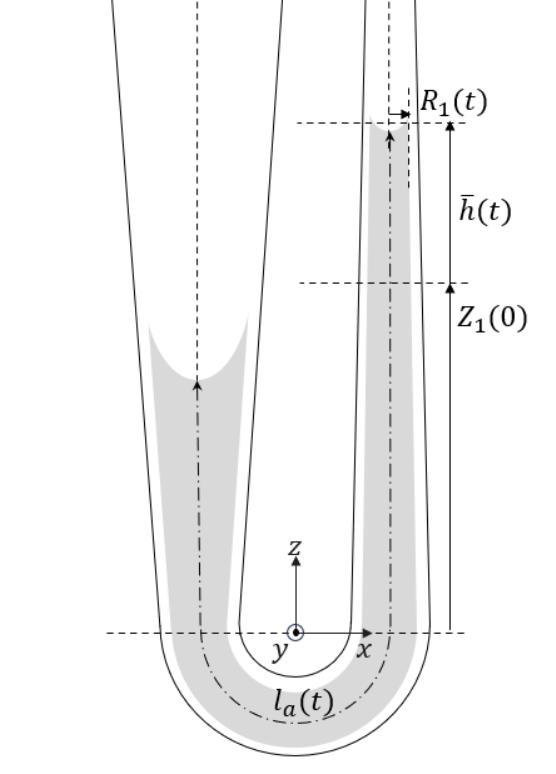}}
 \caption{Schematic of the DUT concept in normal gravity (\ref{subfig:interface1g}) and micro gravity (\ref{subfig:interface0g}) conditions.\label{fig:1}}
\end{figure}

Consider a U-tube characterized by tubes of radii $R_1$ and $R_2$, with $R_1>R_2$, as sketched in Figure~\ref{subfig:interface1g}. The tube is filled by a liquid volume $V_l$ such that the liquid columns on each side are at a level $Z_{1}(0)$ and $Z_{2}(0)$ and the liquid column is of length $l_a(0)$ when this is is subject to normal gravity conditions (Figure~\ref{subfig:interface1g}). 
We assume that the ullage gas in the tube is fully saturated by the liquid vapour; hence, no evaporation occurs at the gas-liquid interfaces. Moreover, let us assume, for simplicity, that the liquid menisci have a spherical shape on both sides of the U-tube. The equilibrium conditions obtained by equalling the hydrostatic pressure to the capillary pressure gives a difference of liquid level $\Delta h_g$ of 

\begin{equation}
\label{surface_T}
\Delta h_g = z_1-z_2= \frac{\Delta p_s}{\rho g}=2 \frac{\sigma \cos(\theta_s)}{\rho a_z} \biggl(\frac{1}{R_1}-\frac{1}{R_2}\biggr)\,,
\end{equation} where $\Delta p_s$ is the capillary pressure, $\theta_s$ is the static contact angle, $\sigma$ is the surface tension at the gas-liquid interface, $\rho$ is the liquid density and $a_z$ is the vertical acceleration. This equals the gravitational acceleration ($a_z=g$) on earth, with $g=9.8$m/s$^2$. As we shall discuss further in the next section, for the conditions considered in this work, the offset $\Delta h_g$ is smaller than $R_2$ and $R_1$.

We are here interested in the motion of the liquid column occurring in the case of a step-like reduction of the gravitational (vertical) acceleration. If this is (significantly) reduced, the hydro-static pressure no longer opposes the capillary forces, and the liquid is set into motion as illustrated in Fig.~\ref{subfig:interface0g}). Because of the different radii in the tubes, mass conservation requires the liquid interface on the smaller radii to move faster than the interface in the larger radii, and the length of the liquid column changes as the liquid moves.

To further promote the capillary-driven displacement of the liquid column and larger contribution of inertia, we consider a Divergent U Tube (DUT) geometry, with converging sections on the two sides with respect to a curvilinear coordinate reference system, as shown in Figure~\ref{fig:1}. Once the liquid column begins its motion, the difference in radii increases, and so does the capillary pressure sustaining the motion. Moreover, the converging section induces a velocity gradient in the stream-wise direction even in the case of constant rising velocity and constant interface shape, thus introducing a non-uniform dynamic pressure near the gas-liquid interface. The radii are linearly varying from the top to the bottom, where they meet via a uniform radius U bending section. 

To characterize the displacement of the liquid column, we focus on the motion of the liquid on the thinner tube, particularly in terms of spatially averaged height: 
\begin{equation}
    \bar{h}\left(t\right)=\frac{1}{\pi R^2_1(t)}\int_{0}^{R}{2\pi\ h\left(r,t\right)rdr}\label{eq:h_average}\,
\end{equation} where $h(r,t)$ is the curve describing the liquid interface with respect to the initial condition $Z_1(0)$, so that one has $\bar{h}(0)=0$ if the shape of the interface is the same at the onset of the motion.

At the limit for vanishing gravitational force, only viscosity and inertia oppose the liquid motion. Let us now consider two extreme scenarios.

If viscosity dominates over inertia, one might expect the flow to reach a fully developed condition with a fully established velocity profile at each cross-stream section and viscous losses equalling the driving capillary pressure. To provide an order of magnitude of the viscous-capillary velocity, let us assume that the flow remains laminar and steady. Hence, we might assume that the pressure drop due to viscous losses obeys the classic Poiseuille law:
\begin{equation}
\label{visc}
\Delta p_v\sim \frac{8 \mu l_{eq}}{R^2_{eq}} U_{eq}\,\rightarrow U_{eq}\sim \frac{\Delta p_s R^2_{eq}}{8 \mu l_{eq}}
\end{equation} where $R_{eq}$ is a sort of hydraulically equivalent radius between the two columns, $l_{eq}$ is an equivalent length of the liquid column, $\mu$ is the dynamic viscosity of the liquid and $U_{eq}$ is the average velocity of the flow.
This velocity could be linked to the contact line velocity $U_{CL}$ from the evolution of the interface shape and the continuity equation.
This formulation is clearly over-simplified and hides the problem complexity in the spatially varying radii and column length in $l_{eq}$ and $R_{eq}$, but allowed for estimating the order of magnitude of the relevant quantities in various scenarios.

If inertia dominates over viscosity, one might expect the liquid column to never reach a stationary velocity during the rise but rather accelerate and eventually over-shoot the equilibrium position. To provide an order of magnitude for the inertial-capillary acceleration, let us assume that the liquid column moves with a constant equivalent acceleration $a_{eq}$ so that the inertial pressure drop $\Delta p_i$ leads to  
\begin{equation}
\label{inertial}
\Delta p_i\sim \frac{V_l \rho a_{eq}}{\pi R^2_{eq}} \rightarrow a_{eq}\sim \frac{\Delta p_s \pi R^2_{eq}}{\rho V_l}\,,
\end{equation} where, once again, the equivalent radius $R_{eq}$ hides the problem complexity in favour of rough estimation.

Assessing a priori which of the two regimes occurs is difficult because it is difficult to estimate (1) what equivalent velocity or accelerations would be observed in the liquid and (2) what contact angle in equation \eqref{surface_T} should be used in dynamic conditions.

This work analyzed this flow configuration to address these points and provide valuable data to validate theoretical and numerical models of dynamic wetting by providing detailed information on the dynamics of the liquid interface, the (dynamic) contact angle evolution, and the flow field in its vicinity.

\section{Methodology}\label{sec:methods}

The experimental set-up and conditions are described in section \ref{SetUP} while section \ref{Meas_tec} reports on the measurement techniques implemented to characterize the interface shape, the liquid column motion and the flow field in its proximity.

\subsection{Experimental configuration and conditions}\label{SetUP}

\begin{figure}[ht!]
       \centering
        \includegraphics[width=0.36\textwidth]{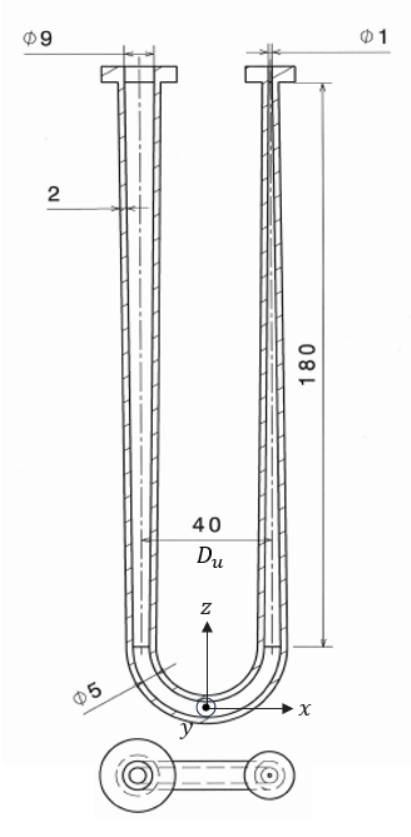}
 \caption{Dimensions (in millimetres) of the DUT geometry considered in this work.}
 \label{fig:2}
\end{figure}

A picture of the DUT geometry used in this work is shown in Figure~\ref{fig:2}. We considered two liquids with significantly different properties, namely HFE7200 and Di-Propylene Glycol (DPG). These are reported in Table~\ref{tab:properties}. The first is a cryogenic propellant replacement fluid characterized by low viscosity, low surface tension, and small static contact angle. The second is a highly viscous fluid. Liquids with similar properties have been extensively used in the literature of dynamic wetting (see, for example, \citet{charlier2022water,IVANOVA2021101399,baumgartner2022marangoni}). 

\begin{table*}
\centering
\begin{tabular}{ccccc}
& \bf density $(\rho)$ & \bf dynamic viscosity $(\mu)$ & \bf surface tension $(\sigma)$ & \bf static contact angle $(\theta_S)$  \\

\bf HFE7200 & 1423.00 $\mbox{kg/m}^3$        & 0.64   $\mbox{mPa}\cdot \mbox{s}$    &  $13.62$ $\mbox{mN/m}$ & $12^{\circ}$\\

\bf DPG & 1022.18 $\mbox{kg/m}^3$        & 62.54   $\mbox{mPa}\cdot \mbox{s}$    &  $34.48$ $\mbox{mN/m}$ & $22^{\circ}$  \\
\end{tabular}
\caption{Properties of the liquids at the experiment conditions (293K). The liquid properties are computed from \citet{rausch2015density} tables for HFE7200 and from \citet{fendu2013vapor} for DPG. Static contact angle values are obtained from the direct measurement of the experimental interfaces before any movement is initiated.}\label{tab:properties}
\end{table*}

\begin{figure*}
    \centering
    \includegraphics[width=0.86\textwidth, trim = 0cm 0cm 0cm 0cm , clip]{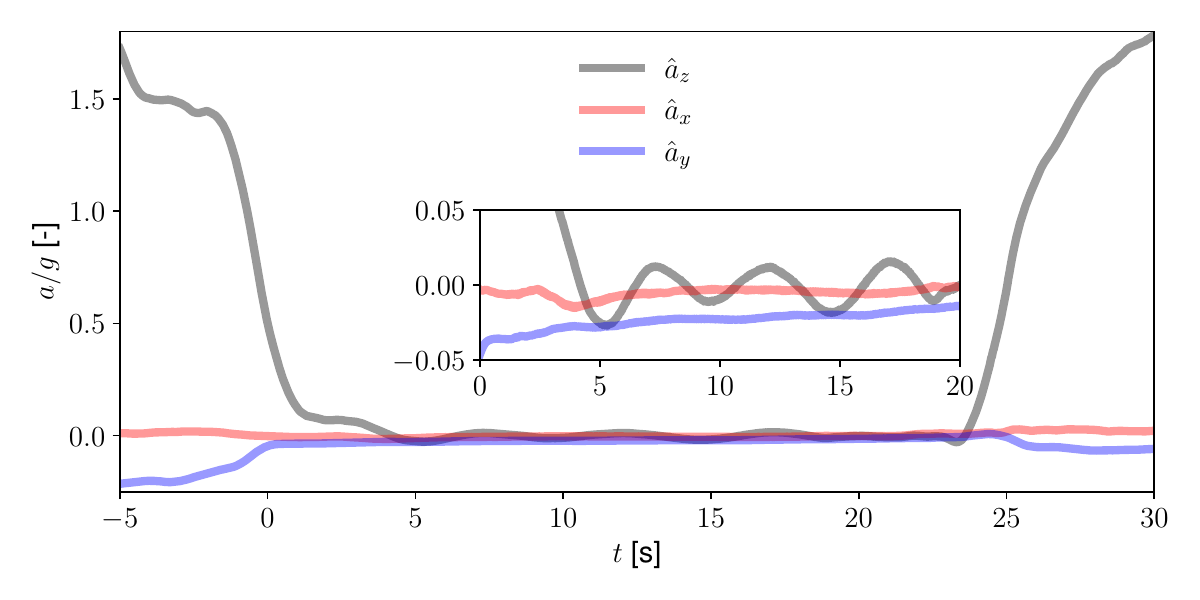}
    \caption{Time series of the accelerations along three axes as measured during one experimental run (\#21) in the 78th ESA parabolic flight campaign. The accelerations are scaled by $g=9.8$m/s$^2$. The three components ($a_x$,$a_y$,$a_z$) are oriented along the axes in Figure \ref{fig:2} with respect to the DUT geometry.}
    \label{Fig3}
\end{figure*}

\begin{table*}[]
\begin{tabular}{lcccccccccc}
                                   & \bf$V_{l}$ (ml)&\bf $R_{1}(0)$ (mm) &\bf $R_{2}(0)$ (mm)& \bf$Z_{1}(0)$ (mm)&\bf $Z_{2}(0)$  (mm) & \bf$\Delta h_0'$ (mm)  & \bf $l_{eq}$ (mm)  &\bf $\Delta P_S$ (Pa)&\bf $U_{visc}$ (mm/s) &\bf $a_{eq}$ (mm/s$^2$) \\ 
\bf HFE7200 \#21 & 2.7 & 2.1 & 2.9 & 37.5 & 36.5        & 1418      & 143              & 3.3         & 28    & 17     \\ 
\bf HFE7200 \#24 & 2.7 & 2.1 & 2.9 & 37.5 & 36.5        & 1418      & 143              & 3.3         & 28    & 17    \\ 
\bf HFE7200 \#26 & 3.0 & 2.0 & 3.0 & 45.0 & 44.2        & 1387      & 164              & 4.1         & 31    & 19    \\ 
\bf HFE7200 \#1.2 & 2.7 & 2.1 & 2.9 & 37.5 & 36.5        & 1418      & 143              & 3.3         & 28    & 17   \\ 
\bf DPG \#21     & 5.2 & 1.4 & 3.6 & 95.1 & 96.8        & 3356      & 267             & 21.5         & 1     & 79     \\ 
\bf DPG \#22    & 5.2 & 1.4 & 3.6 & 95.1 & 96.8        & 3356      & 267              & 21.5        & 1      & 79   \\ 
\bf DPG \#26     & 5.2 & 1.4 & 3.6 & 95.1 & 96.8        & 3356      & 267              & 21.5         & 1    & 79     \\ 
\end{tabular}
\caption{From left to right, the volume of liquid employed in each of the experiments presented, the starting radius of each of the two interfaces and the starting position. $\Delta h_0'$ denotes the theoretical level difference for $a_z=0.0004g$ and $l_{eq}$ denotes the time-averaged liquid column length.\label{tab:cond}}
\end{table*}

The experiments were carried out during the 78th European Space Agency (ESA) parabolic flight campaign. These provide approximately 20 seconds of microgravity, in which the vertical acceleration varies between $10^{-2}g$ and $10^{-6}g$), preceded and followed by approximately 20 seconds of hyper-gravity, in which $a_z\approx1.8g$.

Figure \ref{Fig3} shows an example of acceleration time series measured during the flight along the three axes illustrated in Figure \ref{fig:2}. The data acquisition for each experiment $\#i$ starts with the announcement of the plane being at $40^{\circ}$ inclination. The time axis was arbitrarily set to have $t=0$ when $a_z=0.5g$, which was observed to provide the onset for the liquid motion in most experiments. 

The zoomed view in the region of reduced gravity shows that the acceleration in the vertical direction $z$ oscillates between $a_z=\pm0.01g$, hence repeatedly passing through instants of zero acceleration. The average vertical acceleration within the time window $[10-20]s$ is $a_z=-0.0004 g$. These conditions are generally referred to as microgravity by ESA and in the literature of experiments carried out in parabolic flights.

Concerning the accelerations in the other two components, it is worth noticing that the most disturbing (i.e. largest, especially during the hyper-gravity phase) is along the $y$ direction. Nevertheless, this is orthogonal to the plane along which the center-line of the liquid column moves (cf. Figure \ref{fig:2}). Given the small diameters of the DUT geometry, it is believed that the liquid sloshing induced by this undesired acceleration has a negligible impact on the proposed experiment. 

Since one flight consists of 30 parabolas, experiments were carried out with both liquids using different liquid volumes, hence different initial liquid levels and initial radii. We here report only on the result of seven selected parabolas. Table \ref{tab:cond} collects the relevant information on the experimental conditions for the parabolas considered in this work. These are, from left to right: the liquid volume $V_{l}$ (in ml), the radii $R_1(0)$ and $R_2(0)$ on the two sides of the tube and the associated liquid levels $Z_1(0)$ and $Z_2(0)$ (cf. Figure \ref{subfig:interface1g})
at time $t=0$ (with $a_z=0.5g$), the equilibrium offset $\Delta h'_g$ in the reduced gravity conditions (with $a_z=0.0004g$), computed from \eqref{surface_T}, an estimate for the equivalent length $l_{eq}$, the capillary pressure $\Delta p_s(0)$ defined in \eqref{surface_T}, the equilibrium velocity $U_{eq}$ defined in \eqref{visc} and the equivalent acceleration $a_{eq}$ defined in \eqref{inertial}. The equivalent length and radius were computed by taking an average from the tube geometry, in the case the liquid column moves from its initial condition until the end of the tube on the side with the smallest radius. The value of $R_{eq}=2.5$ mm was considered reasonable for all conditions.

At time $t=0$ ($a_z=0.5g$), the equilibrium offset $\Delta h_g$ in \eqref{surface_T} is of the order of $1.2$ mm for all experiments with DPG and $0.13$ mm for all experiments in HFE7200. Within a time scale of approximately $\approx 5$ s (see Figure \ref{Fig3}), this increases by four orders of magnitude. The equivalent velocities and accelerations in the last two columns of table \ref{tab:cond} provide the order of magnitudes in the extreme cases of negligible inertia and negligible viscosity. Besides the important simplifications mentioned in the previous section, it is worth stressing that these quantities were computed assuming that the contact angle in \eqref{surface_T} remains equal to $\theta_S$, which is clearly an over-simplification. Nevertheless, these estimates also allows for estimating the time required for the liquid column to cover the maximum distance before spilling out of the tubes. This information was important to design the DUT and to define the liquid volume for experiments with DPG and HFE7200.

The liquid volume is a key variable because it sets the initial conditions in terms of the difference of radii on the two sides of the tube (hence the magnitude of the initial capillary pressure) as well as the equivalent column length (hence the magnitude of the viscous losses). Initially, this parameter was adjusted to balance the need for a capillary pressure high enough to initiate liquid motion with the need to capture as much of the liquid's movement on camera as possible. Each flight provides thirty parabolas, organized in six sets of five parabolas separated by a five-minute break. During these breaks, the liquid volume incrementally increased while the experiment repeatability was monitored for each set. The experiments reported in this article are the ones that provided the largest repeatability.

\subsection{Measurement Techniques}\label{Meas_tec}

\begin{figure*}[ht!]
       \centering
        \includegraphics[width=0.9\textwidth, trim = 0.5cm 0cm 0.5cm 0cm, clip]{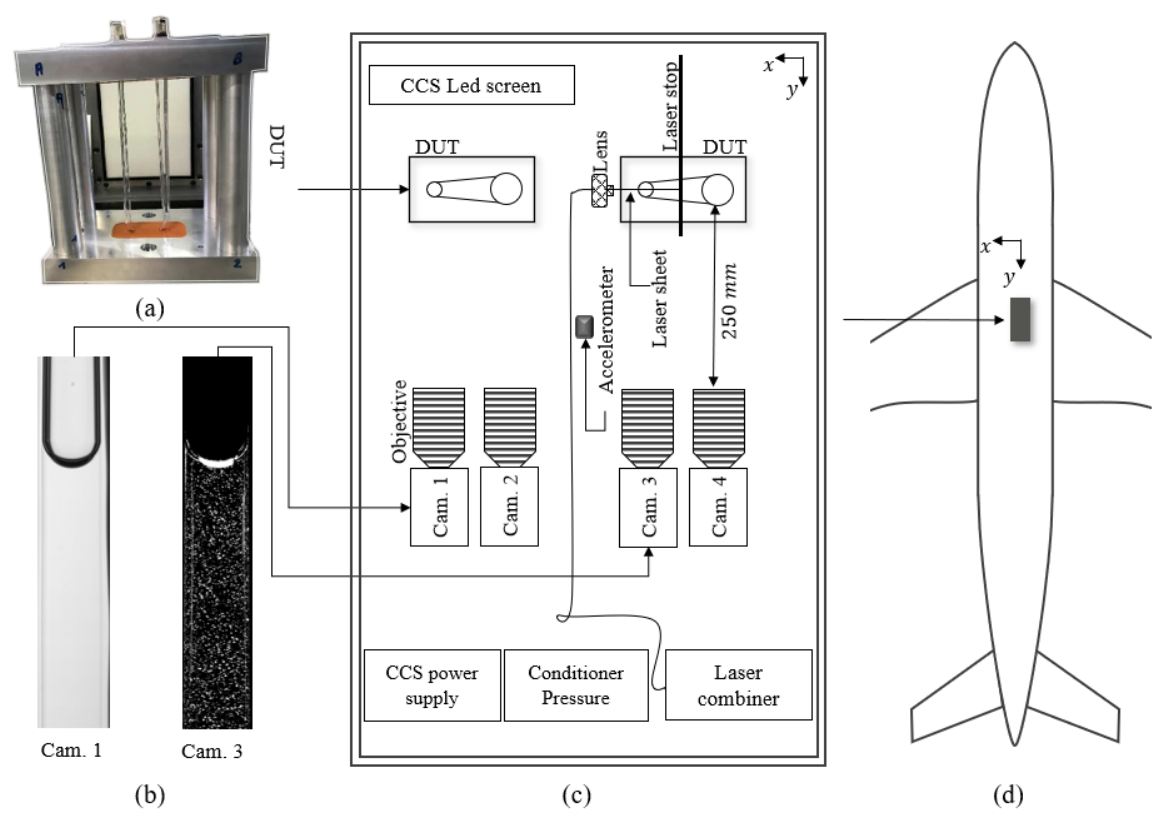}
 \caption{(a) Picture of the DUT. (b) Example images for back-lighting (Cam. 1) and PTV (Cam. 3). (c) Installation on the parabolic flight breadboard. On the left, back-lighting configuration. On the right side the optical set-up for the PTV. (d) Position and orientation of the experimental breadboard (pointed by an arrow) in the plane.}
 \label{fig:setup}
 \vspace*{-3mm}
\end{figure*}

Figure \ref{fig:setup} illustrates the experimental set-up and its installation on the plane. A picture of the DUT configuration is shown in Figure~\ref{fig:setup}a) together with its solid support. Two DUTs were installed in the breadboard as shown in Figure~\ref{fig:setup}c. This allowed for maximizing the number of experimental runs while simultaneously deploying two image acquisition modes, namely back-lighting and particle image velocimetry. Figure~\ref{fig:setup}b shows two sample images from the recording in these two modes. Finally, Figure \ref{fig:setup}d shows the orientation of the breadboard (hence the DUTs) along the plane axes.

The DUTs ullage volume was closed by a flexible connection (not shown in Figure~\ref{fig:setup}), which allows for achieving fully saturated conditions before the experiment begins. Monitoring the over-pressure in the ullage gas and assessing that this equals the saturation pressure of the fluid at the corresponding temperatures allowed for verifying that no evaporation occurs. Four cameras acquired grey-scale images at 300 fps on each side of the two tubes in the two parallel experiments. All cameras mount objectives with 105mm focal lenses and are positioned to acquire the entire gas-liquid interface while spanning the largest possible tube length. The scaling factor of all images is $76.9$ pixel/mm.

Concerning the two visualization techniques, the back-lighting was carried out using LED screens behind the DUTs. This technique was used to characterize the shape of the gas-liquid interface with the highest possible resolution during the full capillary rise. The image processing for identifying the gas-liquid interface is the same employed in \citet{fiorini2022effect} and provides samples of the gas-liquid interface $h(\eta_r,t)$ on a set of points $\eta_r$, with $r\in[1,n_k]$. These were then regressed using Support Vector Regression (SVR) \cite{platt1999probabilistic, Smola2004} to derive a continuous and smooth representation of the gas-liquid interface, facilitating the extraction of the dynamic contact angle formed at the solid surface. Moreover, the SVR allows to precisely define a spatial scale within which the contact angle is reported in this work.

The SVR is a kernel method defining the interface as a linear combination of kernel functions centred in the sample points, i.e.: 
\begin{equation}
\label{SVR}
    h'(\eta,t)=\sum_r\kappa(\eta_r,\eta)\alpha_r(t)\,,
\end{equation} where $\alpha_r(t)$ is the vector of coefficient in each image (to be identified by the regression) and $\kappa(\eta_r,\eta)$ is the kernel function centered at $\eta_r$.

In this work, we use Gaussian Radial Basis Function (RBF) kernels  defined as:
\begin{equation}
    \kappa(\eta_r,\eta)=e^{-\gamma(\eta-\eta_r)^2},
\end{equation} where the scale parameter $\gamma$ controls the length scale of the regression. Defining this as the distance within which $\kappa=0.5$, one has $l_c=\sqrt{\ln(2)/\gamma}$. In this work, we consider $l_c\approx 1/10$ of the local radius of the tube.


The second visualization technique employed during the campaign is Particle Tracking Velocimetry (PTV), with the goal of characterizing the velocity field near the gas-liquid interface on the smaller side of the tube. This acquisition was carried out only for the HFE7200 due to time constraints. The illumination for this technique was provided by a laser sheet on one side of the tube. The liquid was seeded with FMR-1.3 1-5 $\mu m$ Red Fluorescent Microspheres from Cospheric. The impact of the particles on the fluid properties was analyzed on ground in classic capillary-rise experiments and was found to be negligible.

 The recordings are pre-processed by compensating for the vibrations of the camera objective (as in \citet{fiorini2023MST}) and optical distortions due to light refraction (as in \citet{fiorini2022effect}). The PTV images are pre-processed using the POD-based background removal introduced by \citet{mendezpod} while the PTV interrogation was carried out using the Tractrac processing tool \citep{heyman2019tractrac}. A dynamic Region of Interest (DROI) for the processing was identified by performing a gradient-based edge detection on the POD-filtered images to detect the gas-liquid interface and computing the average velocity (see \citet{fiorini2023MST,ratz2023analysis}). Nevertheless, the PTV acquisitions do not allow for precise detection of the gas-liquid interface, which is carried out solely from the recordings in backlight configuration. 

The PTV measurements were enhanced using the constrained Radial Basis Function approach developed by \citet{sperotto2022meshless} and released in the open-source software SPICY \citep{spicy}. The regression was carried out on an ensemble of three consecutive frames to increase the vector density. 


\section{Results}\label{sec:results}
\subsection{Capillary rise}\label{sec:undulation}
\begin{figure*}
\captionsetup[subfigure]{skip=-0.05\textwidth,margin=0.97\textwidth}
\begin{subfigure}[t]{0.99\textwidth}
    \includegraphics[width=\textwidth, trim = 0.3cm 4.5cm 0.3cm 0.5cm , clip]{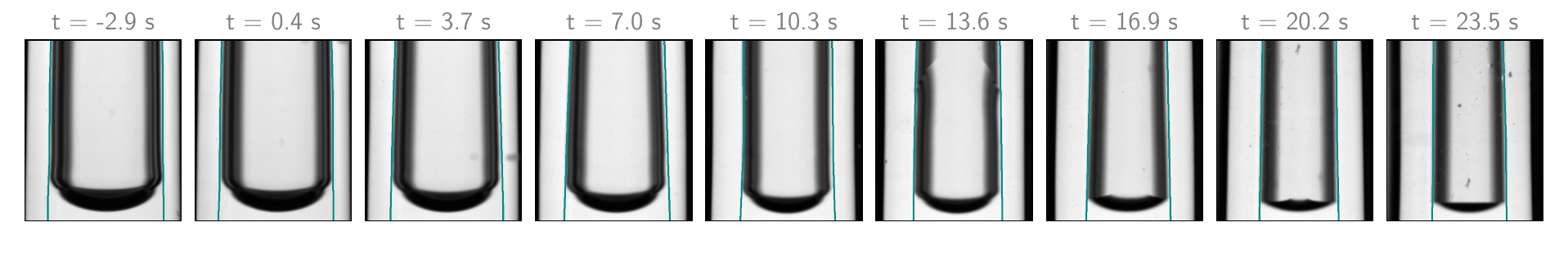} 
    \caption{\label{fig:ExperimentsDiPropyleneGlycol}}
\end{subfigure}
\begin{subfigure}[t]{0.99\textwidth}
    \centering
    \includegraphics[height=6.9cm, trim = 0 0.5cm 0 0.5cm 0, clip]{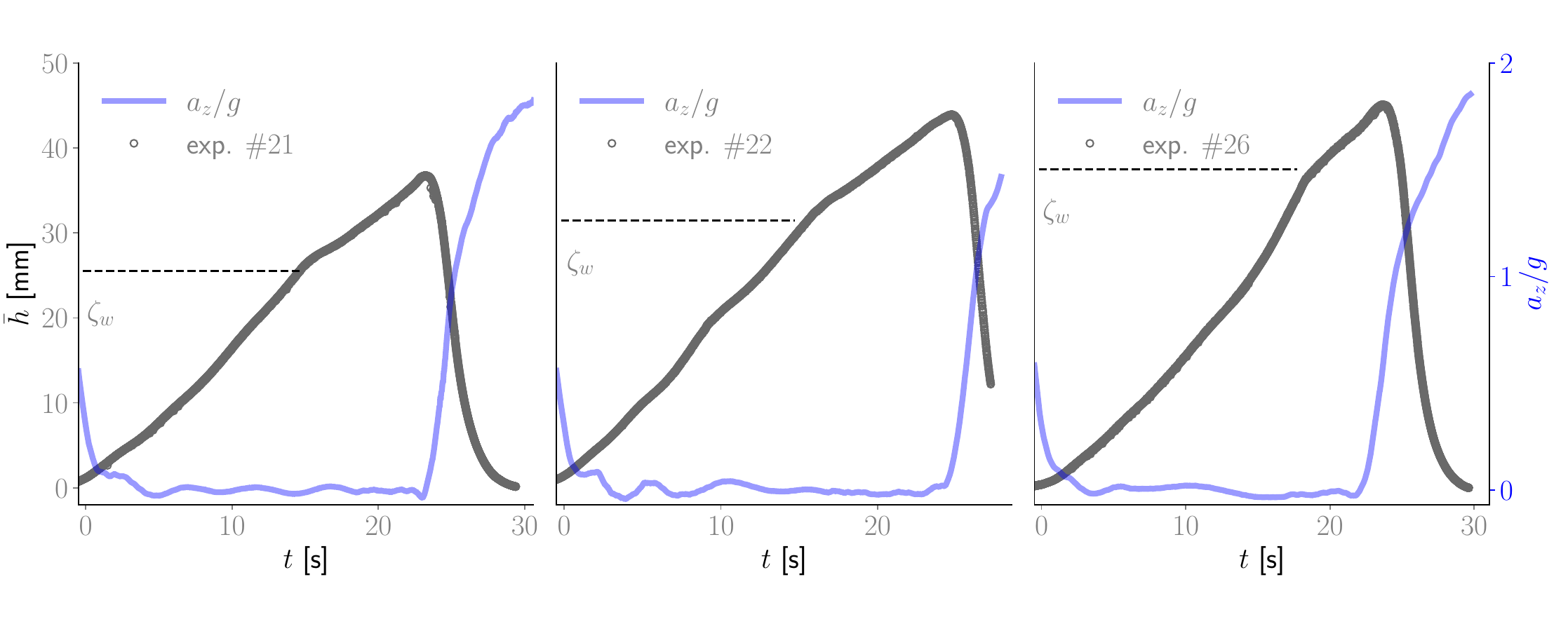} 
    \caption{\label{fig:ComparisonCapillaryRiseDiPropyleneGlycol}}
\end{subfigure}
\caption{Figures on top: snapshots of the interface of DPG during exp. $\#21$. The position of the solid surface is indicated in blue. Bottom, capillary rise experiments with DPG. The dashed line indicates the position where the wetting conditions change. \label{fig:DPGglobal}}
\end{figure*}

Figure~\ref{fig:ExperimentsDiPropyleneGlycol} presents snapshots at several interface positions along the tube's axis recorded during the capillary rise experiments with DPG. Figure~\ref{fig:ComparisonCapillaryRiseDiPropyleneGlycol} shows the capillary rise of DPG for three experiments and the envelope of gravitational acceleration along the z-axis for each experiment, based on the axis convention of Figure~\ref{fig:2}. Table \ref{tab:cond} lists the initial conditions of the interfaces before the start of the capillary rise.
The gas-liquid interface rises slowly along the tube's axis, remaining within the camera's field of view. A sharp reduction of the interface curvature occurs after crossing the position $z = \zeta_w$ identified in the figures. This condition occurs at $t\approx16s$ for exp. $\#21$, at $t\approx17s$ for $\#22$ and at $t\approx19s$ for $\#26$. Crossing this position produces a visible change of the interface shape as illustrated in the snapshots of Figure~\ref{fig:ExperimentsDiPropyleneGlycol} between $t=13.6$s and $t=16.9$s. Figure~\ref{fig:ComparisonCapillaryRiseDiPropyleneGlycol} reveals a change in the capillary rise profile's slope at $z \approx \zeta_{w}$ and a higher final interface position with each rise. Correspondingly, the value of $\zeta_w$ increases over time. This displacement is due to a residual film of DPG covering part of the inner surface of the tube and making the interface advance over a pre-wetted portion of the tube, which increases at each run as shown by the dashed line in Figure~\ref{fig:ComparisonCapillaryRiseDiPropyleneGlycol}. For $z>\zeta_w$, the interface moves over a dry surface with a significant increase in contact angle and reduction in the interface curvature. 

Figure~\ref{fig:ComparisonCapillaryRiseDiPropyleneGlycol} illustrates how the pre-wetted height of the tube increases with the number of repetitions. The snapshot at $t=13.6$s captures the transition between the pre-wetted and dry surfaces. In all three cases, the capillary rise of the interface exhibits an approximately constant velocity of 2 mm/s for the pre-wet surface and 1 mm/s for the dry surface. These velocities are compatible with the $U_{visc}$ computed for the visco-capillary dominated regime, as shown in Table~\ref{tab:cond}. Conversely, the hypothesis of uniform acceleration $a_{eq}$ leads to a much faster capillary rise, inconsistent with the experimental observations. This result suggests that, for the case of DPG, the interface motion is driven by the balance of viscous and capillary forces, conditioned by the wetting status (pre-wet or dry) of the solid surface.

\begin{figure*}
\captionsetup[subfigure]{skip=-0.05\textwidth,margin=0.97\textwidth}
\begin{subfigure}{0.99\textwidth}
        \centering
    \includegraphics[width=\textwidth, trim = 0.3cm 4.2cm 0.3cm 0.5cm , clip]{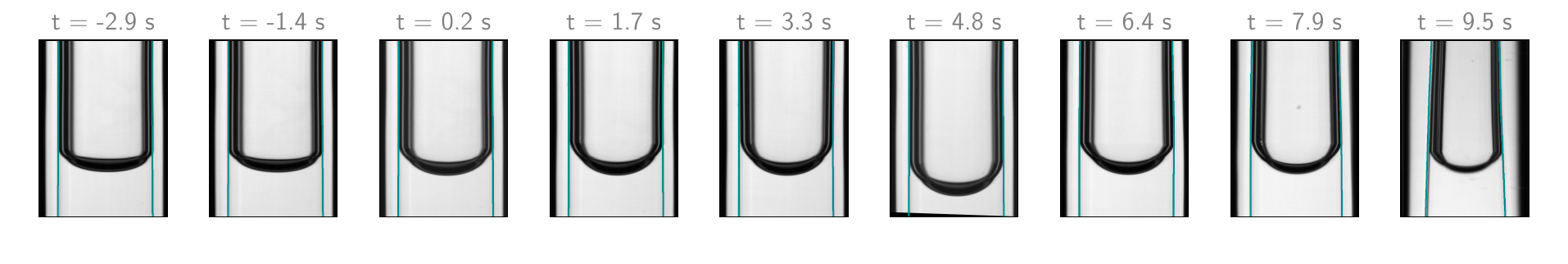} 
    \caption{\label{fig:ExperimentsHFE7200}}  
\end{subfigure}
\begin{subfigure}{0.99\textwidth}
    \centering
    \includegraphics[height=6.9cm, trim = 0 0.5cm 0 0.5cm 0, clip]{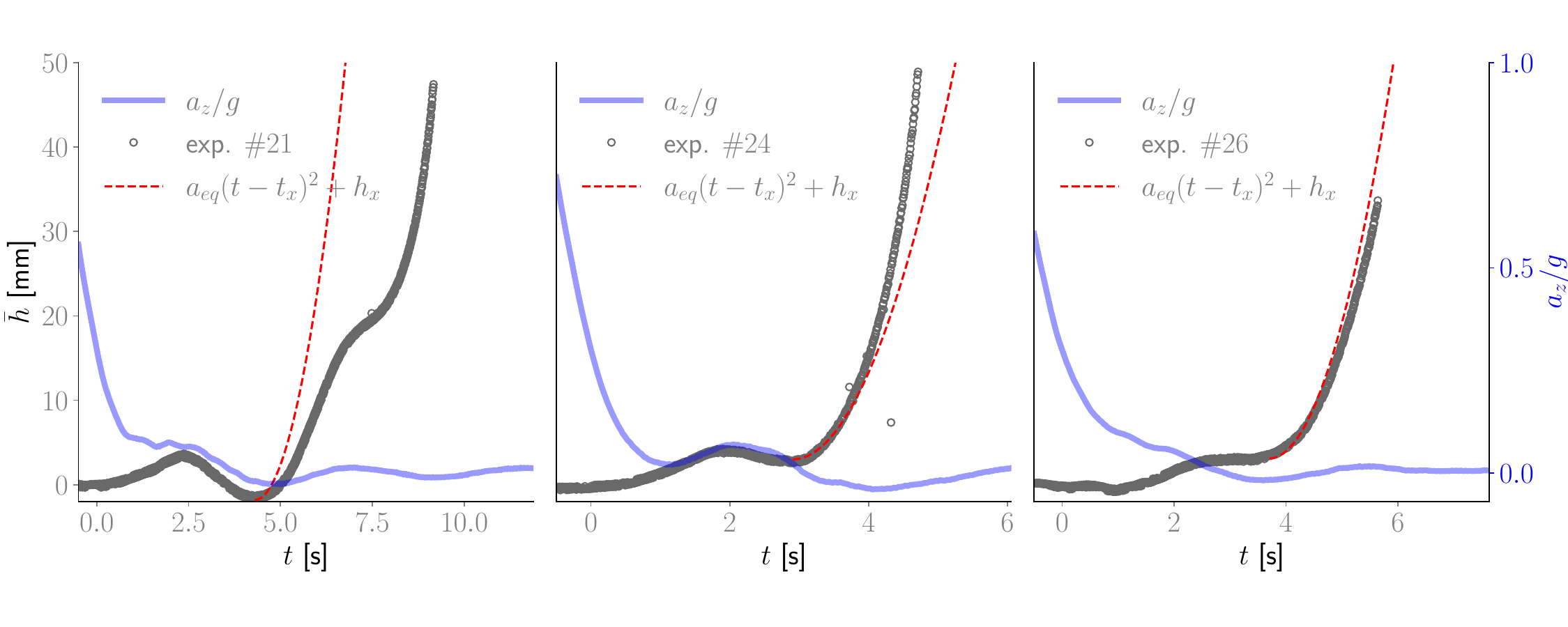} 
    \caption{\label{fig:ComparisonCapillaryRiseHFE7200}}
    \end{subfigure}
    \caption{On top, snapshots of the interface of HFE7200 during exp. $\#21$. The position of the solid surface is indicated in blue. Bottom, capillary rise of HFE7200 for different profiles of vertical acceleration. Exp. $\#21$ shows also a deceleration at $\approx 7s$ due to gravity level increase.}
    \label{fig:HFE7200global}
\end{figure*}

Figures~\ref{fig:ExperimentsHFE7200} and~\ref{fig:ComparisonCapillaryRiseHFE7200} present the same analysis for experiments conducted with HFE7200. Table \ref{tab:cond} lists the initial condition of the interface also for this set of experiments. In this case, the separation between ``wet" and ``dry" surface regions could not be identified and the wetting behaviour of the fluid suggests a ``pre-wet" solid surface throughout the entire experiment. Comparing to the viscous DPG, Figure~\ref{fig:ComparisonCapillaryRiseHFE7200} illustrates that the capillary rise of HFE7200 is heavily dependent on the vertical acceleration profile, which is more significantly impacted by the flight maneuver. Nevertheless, common features are observed in all experiments. First, as the vertical acceleration decreases, the capillary rise does not start immediately. Instead, the interface exhibits one or more oscillations until gravity levels drop to about $10^{-3}g$. Then, once the capillary rise is initiated, the interface accelerates continuously until it leaves the camera's field of view. 
The oscillating behaviour of the fluid and the absence of a constant rising velocity prevents the comparison with the values of $U_{visc}$ in Table \ref{tab:cond}. However, after the initial oscillations, the three experiments reach a nearly constant acceleration which is kept for a significant portion of the camera's field of view. To illustrate this, we arbitrarily set initial conditions $\mbox{IC}=[t_x,h_x]$ from which the interface exhibits a clear monotonic rise and illustrate in Figure \ref{fig:ComparisonCapillaryRiseHFE7200} the interface trajectory in the case of uniformly accelerated conditions using the viscous-capillary acceleration in \eqref{inertial} (see Table \ref{tab:cond}).

The overall agreement with the experimental interface position suggest that these experiments are in the inertial-capillary dominated condition. The discrepancies and the differences in the observed trends can be explained by the higher sensitivity to the spurious oscillations, which are inevitably different between flights. For example, the interface slows down before accelerating again in experiment \#21, while the acceleration further increases and exceeds the estimate in Table~\ref{tab:cond} in experiment \#24. 
It is worth stressing that these considerations are limited to a short distance after the inception of the rise, since the interface quickly moves away from the field of view. The underestimation of the interface acceleration likely becomes even more pronounced in later stages. Nevertheless its interesting to note that the shape of the interface qualitatively remains unchanged in the initial part of the experiment but flattens towards the end, as shown by the last frames in Figure~\ref{fig:ExperimentsHFE7200}. This is addressed to the inertial effects near the interface, and it is further analyzed with the help of PTV measurements in the following section.

\subsection{Interface shape and contact angle}\label{sec:result_model}
\begin{figure}[h]
\captionsetup[subfigure]{skip=-0.01\textwidth,margin=0.2\textwidth}
    \centering
      \begin{subfigure}{0.49\textwidth}
        \includegraphics[height = 6.5cm,  clip]{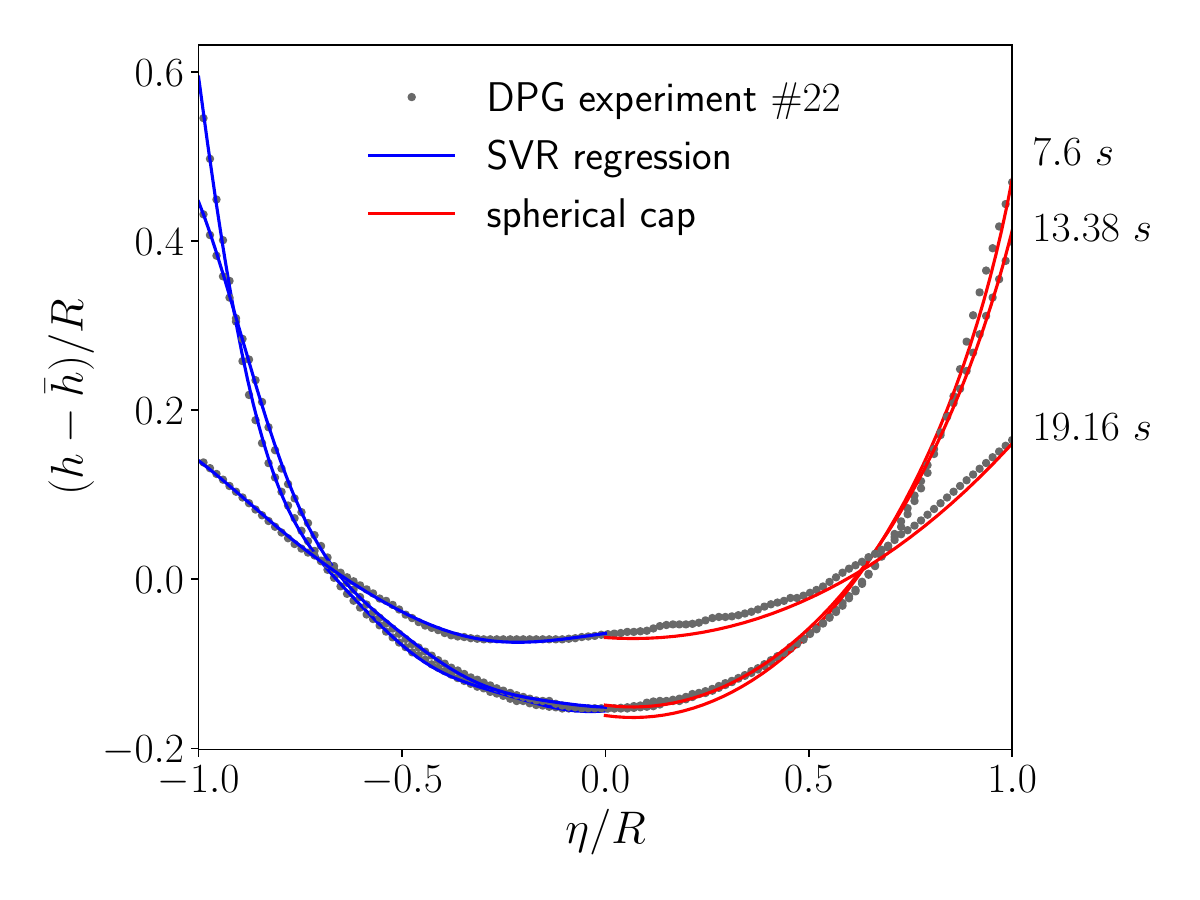}
        \caption{\label{subfig:sphericityDPG} Di-Propylene Glycol}
    \end{subfigure}
    \begin{subfigure}{0.49\textwidth}
        \includegraphics[height = 6.5cm,  clip]{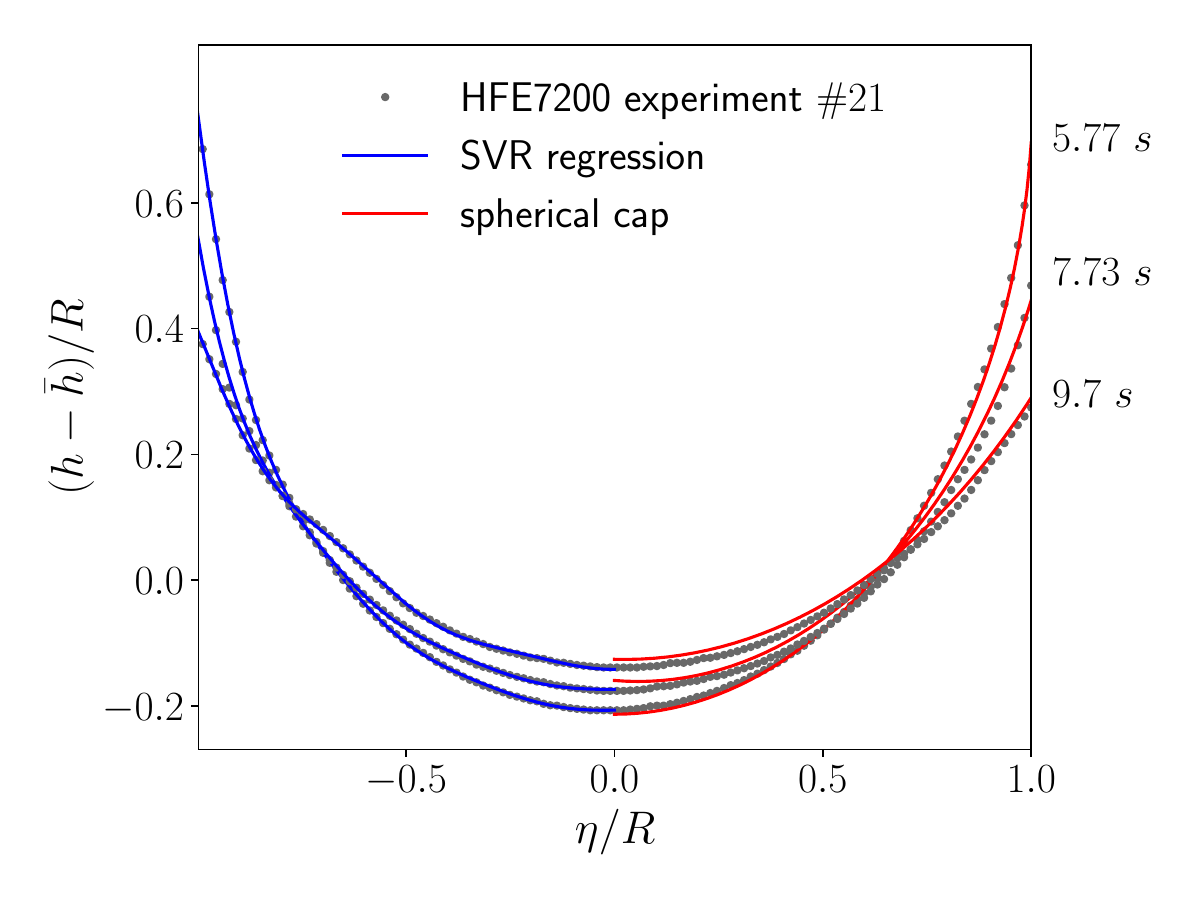}
        \caption{\label{subfig:sphericityHFE} HFE7200}
    \end{subfigure}
    \caption{Comparison of the interfaces with spherical cap (right) and SVR model (left).}
    \label{fig:interfaceShapes}
\end{figure}

Figure~\ref{fig:interfaceShapes} presents a comparison between the interface shapes of DPG (Figure~\ref{subfig:sphericityDPG}) and HFE7200 (Figure~\ref{subfig:sphericityHFE}), along with the results of the SVR in \eqref{SVR} (in blue, on the left) and the assumption of spherical interface (in red, on the right).
Remarkably, within the kernel's length scale of $l_c\approx 0.7\mbox{mm}$ both fluids maintain a spherical shape throughout the entire microgravity phase, even when the interface flattens due to the different dynamics previously discussed for DPG and HFE7200. Therefore, the interface shape can be described given the tube radius and the dynamic contact angle.

\begin{figure*}[]
    \centering
    \includegraphics[width=0.9\textwidth]{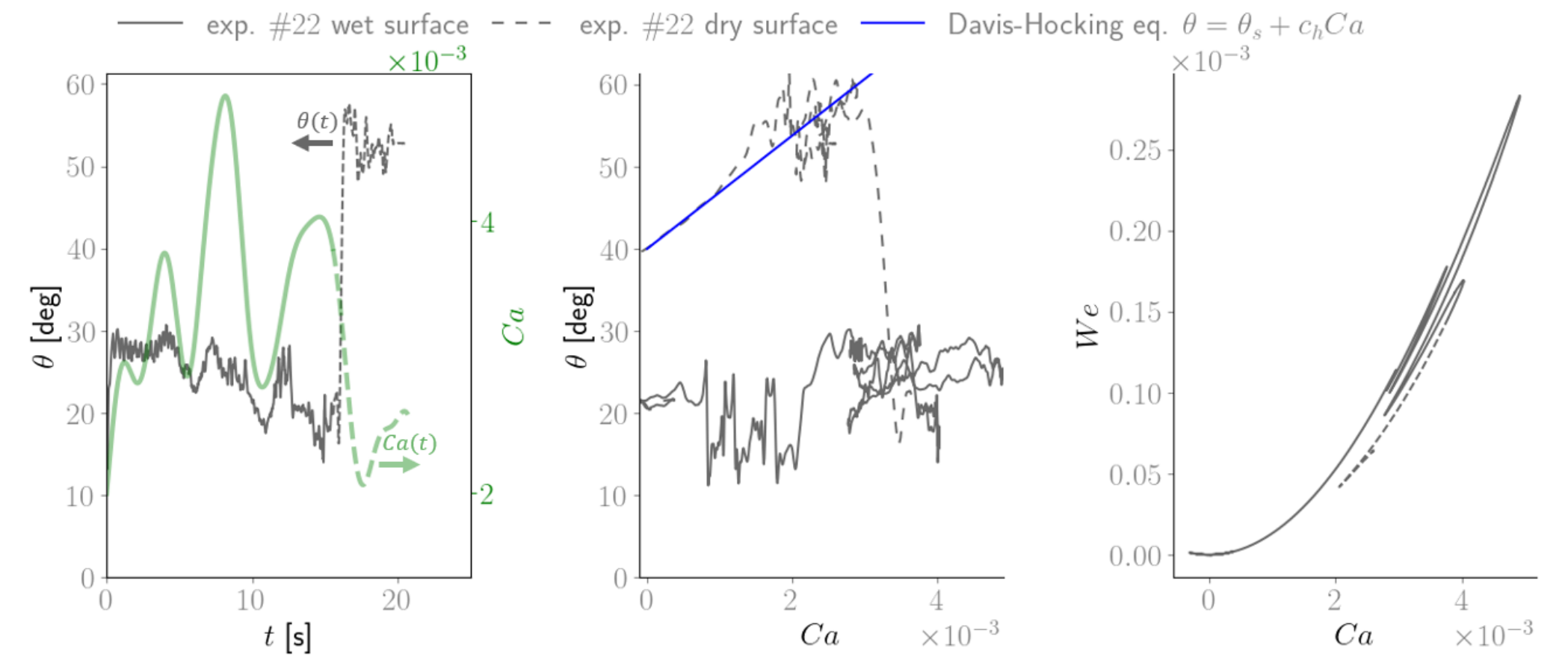}
    \vspace*{-20mm}  
\begin{center}
\hspace{2cm}
    (a)
\hspace{5cm}
    (b)
\hspace{4cm}
    (c)
\end{center}
\vspace*{10mm}  
 \caption{DPG run $\#22$. From left to right: (a) Dynamic contact angle against time. The Capillary number is also shown on the right axis (b) Plot of the measured contact angle against the capillary number. The prediction with Davis-Hocking correlation is also shown for the dry surface plot ($c_h=120$) (c) Comparison of the interface $\mathrm{We}$ against $\mathrm{Ca}$. The Figure distinguishes with different line styles the data belonging to a pre-wet surface to the ones of a dry surface.}
 \label{fig:DPGContactAngle}
\end{figure*}

\begin{figure*}[]
    \centering
    \includegraphics[width=0.9\textwidth]{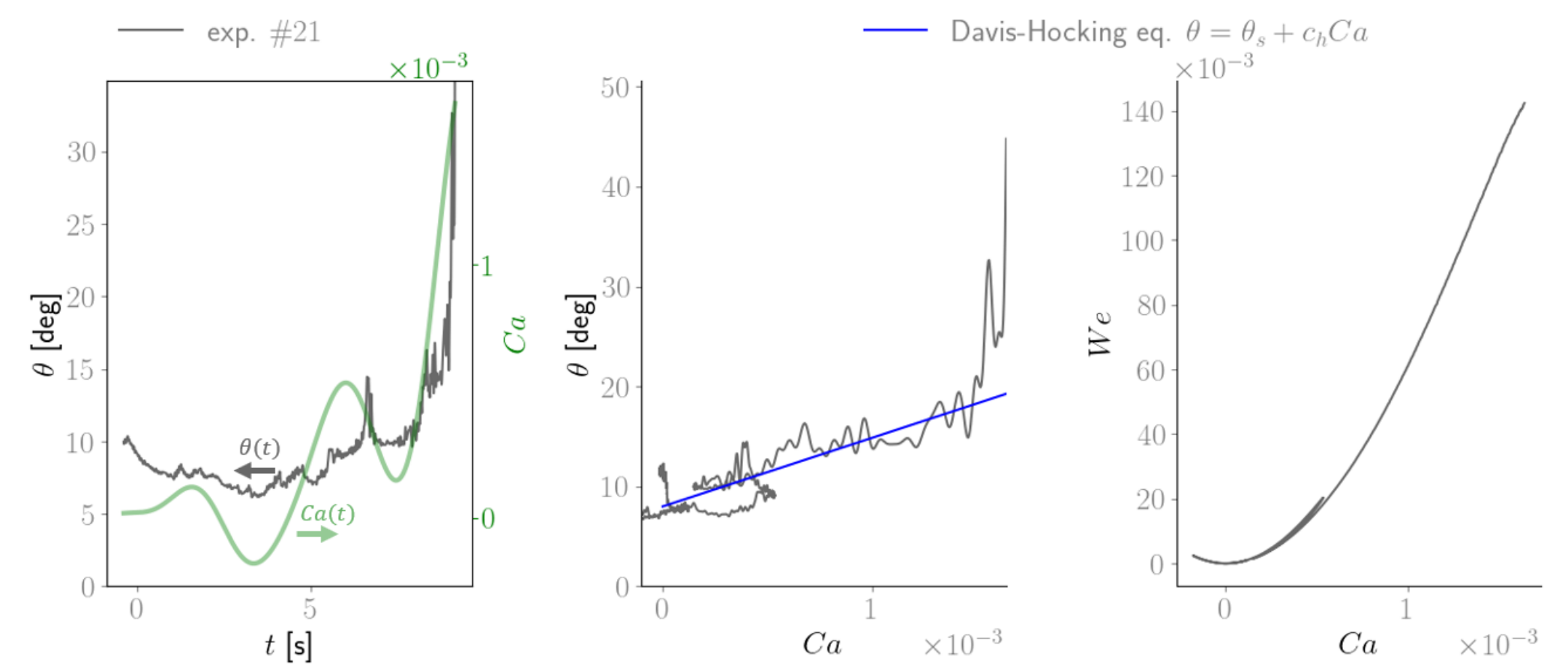}
    \vspace*{-20mm}  
\begin{center}
\hspace{2cm}
    (a)
\hspace{5cm}
    (b)
\hspace{4cm}
    (c)
\end{center}
\vspace*{10mm}  
 \caption{Same as Figure~\ref{fig:DPGContactAngle}, but for runs with HFE7200 (run $\#21$).}
 \label{fig:HFEContactAngle}
\end{figure*}

Figure~\ref{fig:DPGContactAngle} compares the interface contact angle obtained by extrapolating the SVR model to the solid surface with the capillary number for the experiment $\# 21$ with DPG. The other runs exhibit similar behaviour. The experimental points are distinguished with different line styles associated with the different wetting conditions of the tube. In Figure~\ref{fig:DPGContactAngle}a, the transition from the wetted (solid line) to the dry region (dashed line) is evident, accompanied by a sharp increase of the contact angle. Figure~\ref{fig:DPGContactAngle}b shows the contact angle decreasing during the capillary rise on the wetted surface. No clear correlation with the capillary number, i.e. the contact-line velocity, is observed.
On the other hand, in the dry region, the contact angle exhibits a linear relationship with the capillary number, as hypothesized by \citet{davis1980moving} and \citet{hocking1987damping} and shown by the dark blue line in Figure~\ref{fig:DPGContactAngle}b. The same relationship was also observed more recently by \citet{xia2018moving} and \citet{fiorini2023PRF}. Figure~\ref{fig:DPGContactAngle}c plots the Weber number at the interface, computed as $\mathrm{We}=\rho R U_{CL}^2/\sigma$, and the capillary number. The different trajectories for interface motion along "dry" and "pre-wet" surfaces are visible, with the first characterized by a slower interface for the same tube radius. 

The same analysis is presented for the experiment with HFE7200 in Figure~\ref{fig:HFEContactAngle}. Figure~\ref{fig:HFEContactAngle}a shows the dynamic contact angle initially decreases, as in the case of DPG, then increases rapidly at the latest stages of the capillary rise. From a different perspective, Figure~\ref{fig:HFEContactAngle}b shows that the linear relationship between the contact angle and the capillary number changes after $\mathrm{Ca}=1.10^{-3}$ where the Weber number exceeds $\mathrm{We}=50.10^{-3}$. In the case of HFE7200, the rapid reduction of interface curvature is addressed to its sudden acceleration. The next section analyzes this dynamics with the PTV measurements. The plane of $(\mathrm{We},\mathrm{Ca})$ in Figure~\ref{fig:HFEContactAngle}c shows that the experiment with HFE7200 approximately follows the same curve throughout the observation period, given the absence of a transition between "wet" and "dry" regions.

\subsection{Flow dynamics in the vicinity of the interface}\label{sec:result_PTV}
\begin{figure*}[]
    \centering
    \includegraphics[width=0.95\textwidth, trim = 0 0.cm 0cm 0cm, clip]{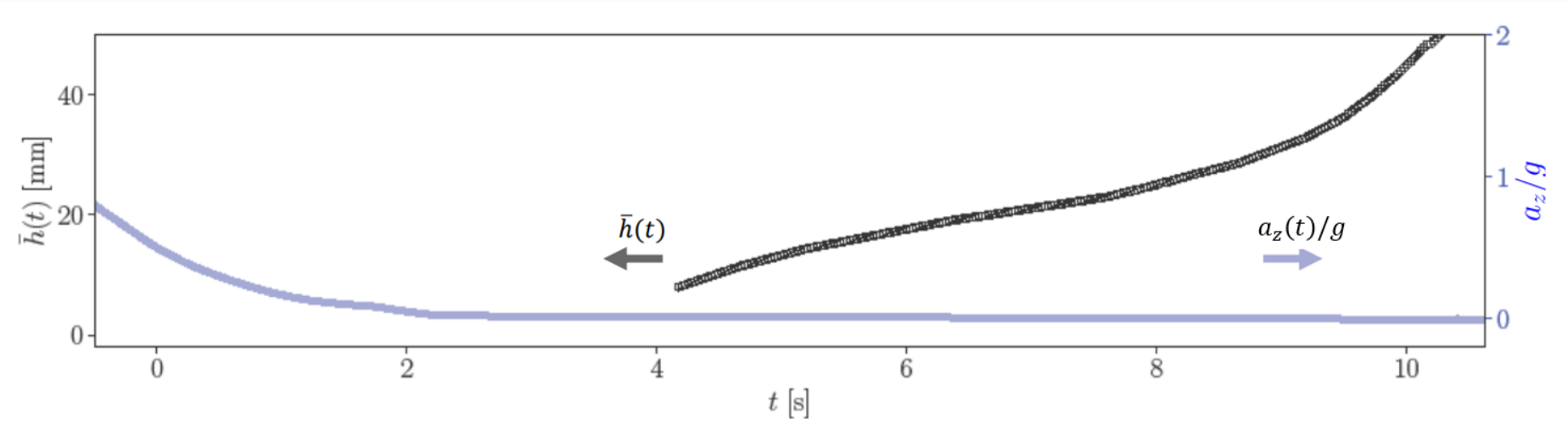}
 \caption{HFE7200 run $\#1.2$. Plot of the interface height (left axis) and the vertical acceleration (right axis) as a function of time. }
 \label{fig:run31height}
\end{figure*}

\begin{figure*}[]
    \centering
    \includegraphics[width=\textwidth, trim = 0cm 0cm 0cm 1cm, clip]{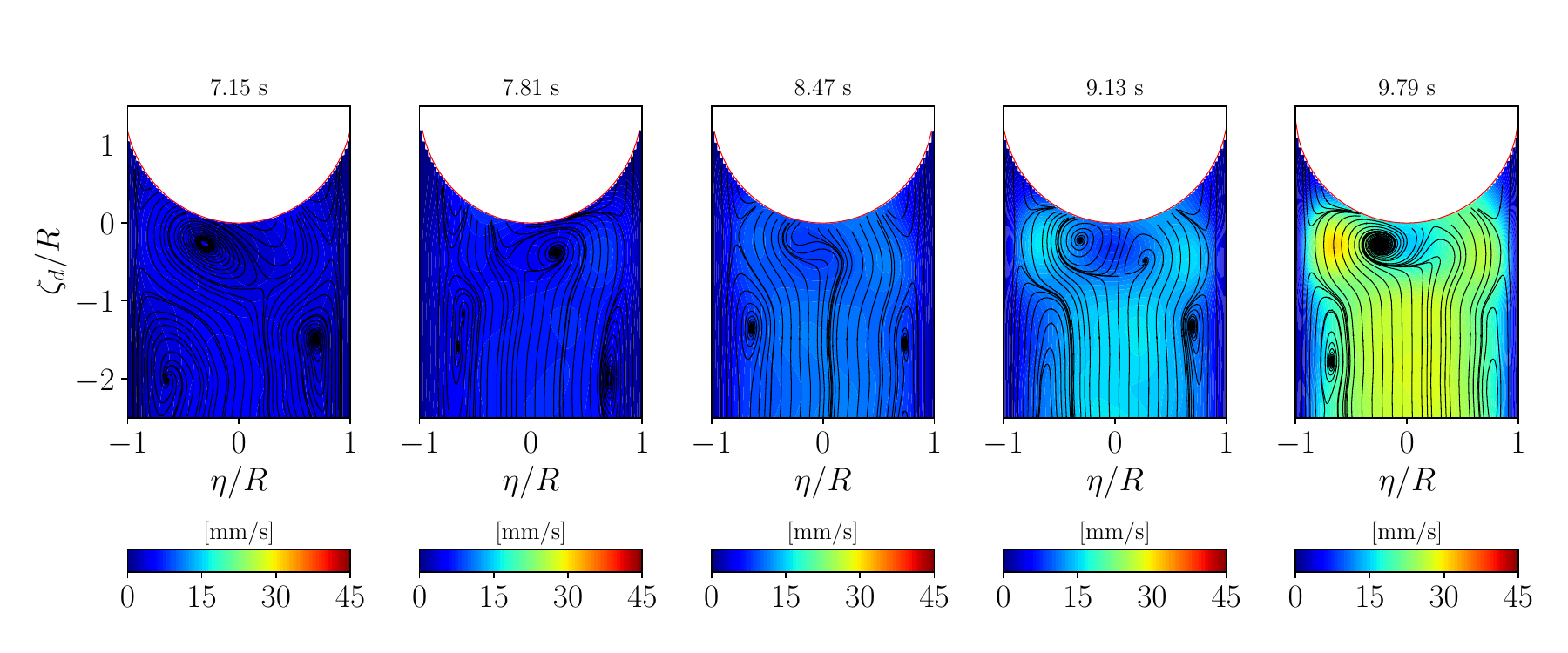}
 \caption{HFE7200 run $\#31$. Contour plot of velocity magnitude in the proximity of the interface. On top, streamlines for a moving reference frame with the velocity of the meniscus.}
 \label{fig:velocity_fields}
\end{figure*}
\begin{figure*}[]
    \centering
    \includegraphics[width=0.95\textwidth, trim = 0 0.cm 0cm 0cm, clip]{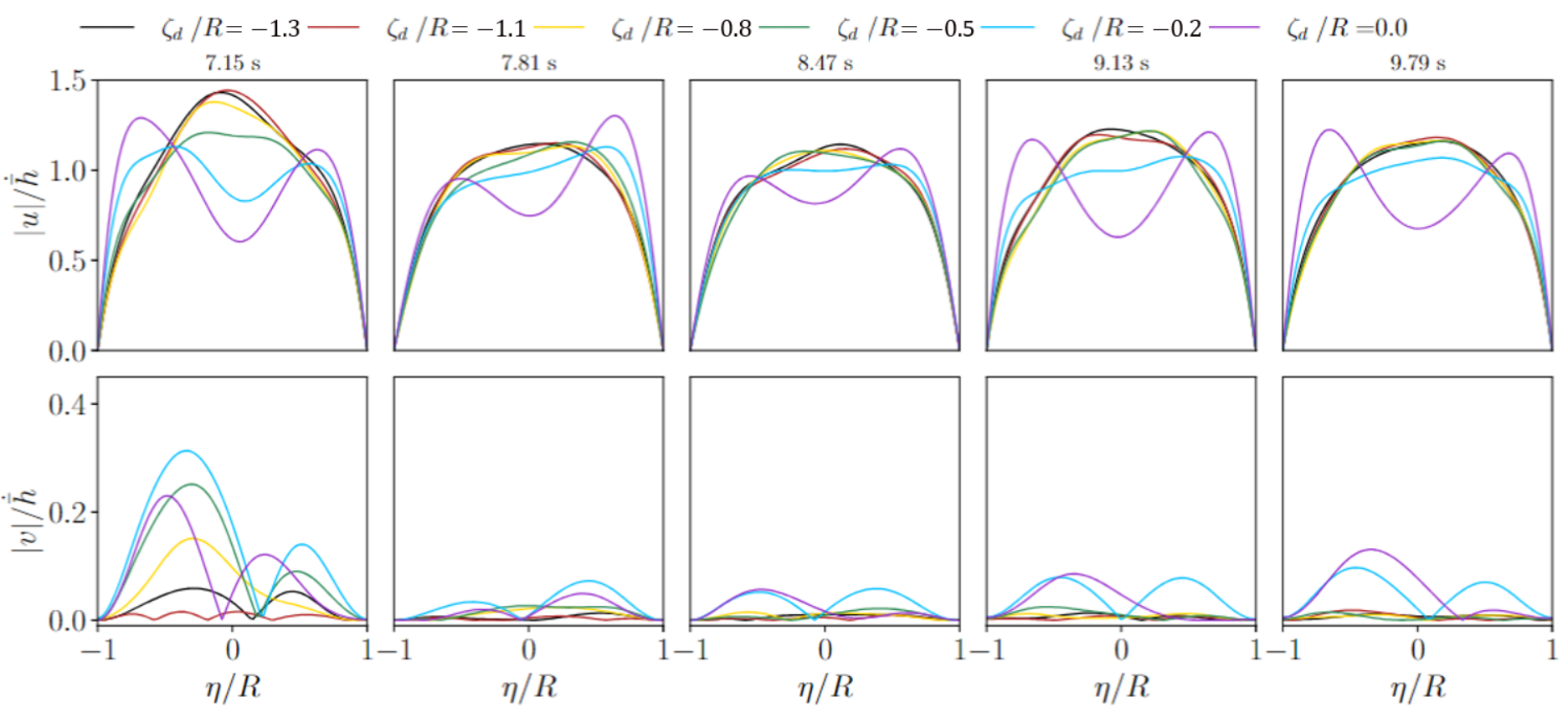}
 \caption{HFE7200 run $\#1.2$. streamwise and radial velocity profiles during the capillary rise.}
 \label{fig:velocity_profiles}
\end{figure*}

\begin{figure*}[]
    \centering
    \includegraphics[width=\textwidth, trim = 0.5cm 0cm 0.5cm 0cm, clip]{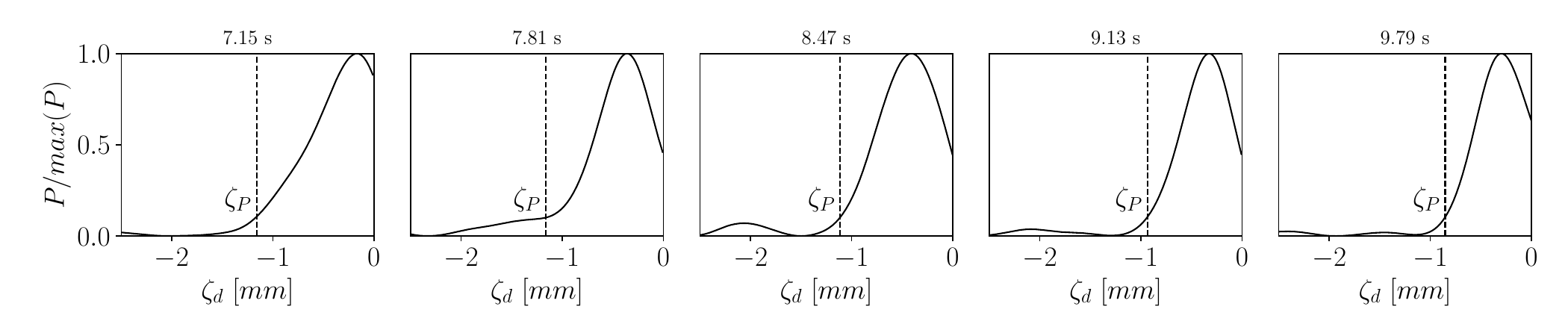}
 \caption{Evolution of the normalized error between the streamwise velocity profile and the theoretical parabolic profile.}
 \label{fig:parabolicity}
\end{figure*}

\begin{figure*}[]
    \centering
\includegraphics[width=0.9\textwidth, trim = 0.5cm 0cm 0.5cm 0cm, clip]{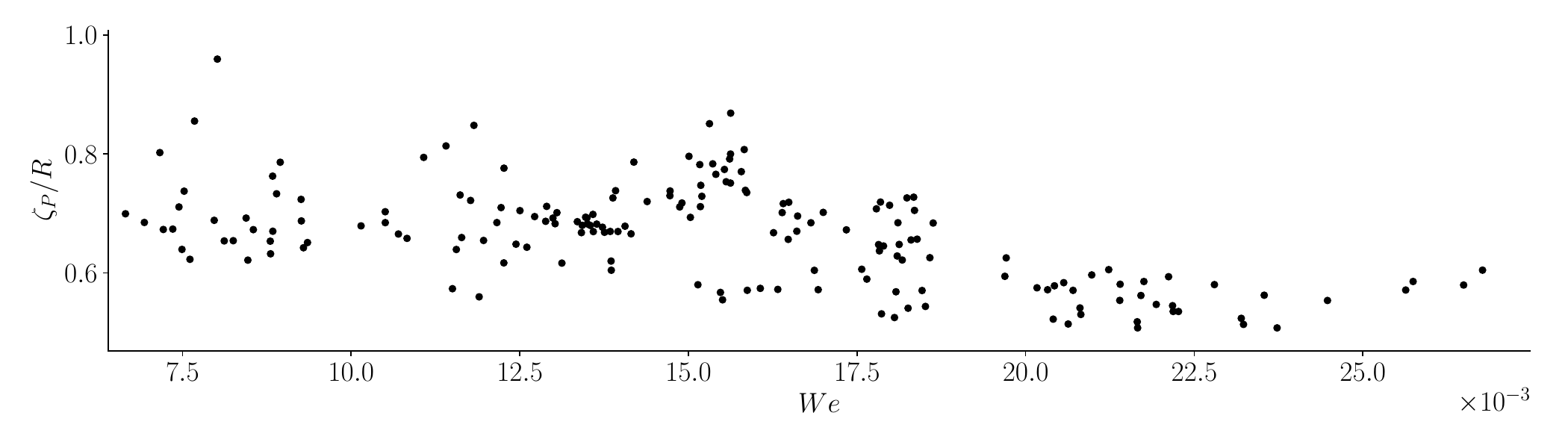}
    \vspace{-0.5cm}
 \caption{Evolution of $\zeta_P/R$ with the interface Weber number.}
 \label{fig:parabolicityAll}
\end{figure*}

To examine the velocity field near the liquid interface, we use the test case introduced by \citet{fiorini2023MST}, where the capillary rise of HFE7200 seeded with fluorescent polymer particles was presented, and a parabolic velocity profile was shown at distances from the interface greater than 2 mm. Figure~\ref{fig:run31height} shows the position of the interface for the considered test case analogously to the experiments illustrated by Figure~\ref{fig:ComparisonCapillaryRiseHFE7200}. In this analysis, we focus on the velocity field within 2 mm below the interface and introduce a reference frame moving with the velocity of the meniscus $\dot{\bar{h}}$ and with $\zeta_d=z-h(0,t)$. Figure~\ref{fig:velocity_fields} displays the velocity fields in this region in terms of absolute magnitude and the flow streamlines computed using the moving reference frame. The red curve represents the detected interface within each image.

Figure~\ref{fig:velocity_fields} illustrates both the presence of counter-rotating vortices underneath the interface, akin to configurations observed by \citet{davis1974motion} and \citet{ratz2023analysis}, and larger re-circulation zones at a further distance. 
In the reference frame moving with the interface, the meniscus is slower than the fluid along the centerline. As a result, the interface behaves as an obstacle: the fluid must decelerate and move towards the contact line, producing the swirling motion depicted by the streamlines in Figure~\ref{fig:velocity_fields}.
Regarding the axial ($\zeta$) and radial ($\eta$) velocity profiles, indicated as $u=u(\eta,\zeta,t)$ and $v=v(\eta,\zeta,t)$, respectively, Figure~\ref{fig:velocity_profiles} illustrates their variation for decreasing distances from the meniscus, $\zeta_d/R$. Both quantities are made dimensionless with respect to the meniscus velocity $\dot{\bar{h}}$. At low interface accelerations ($t<9$ s), the streamwise velocity profile at $\zeta_d/R>-0.8$ from the interface is relatively flat, with the centerline velocity decreasing as it moves towards the interface. At larger accelerations ($t>9$ s), the parabolic profile appears closer to the interface. 
The shape of the velocity profile changes gradually from the parabolic shape at $\zeta_d/R<-1$ to an approximately square profile at $\zeta_d/R\approx-0.5$ and finally accelerating towards the contact line at the bottom of the interface ($\zeta_d/R\approx 0$). For $\zeta_d/R<-1$, the radial velocity component can be considered flat compared to the streamwise velocity. 

Closer to the interface, we observe the flow increasing velocity towards the side of the tube. This is necessary to adapt the shape of the streamwise velocity profile to the uniform velocity of the meniscus. A local maximum of the profile can be localized at the positions $|\eta|/R\approx0.5$. However, the radial profile tends to lose axial symmetry at short distances from the interface. At the latest stage of the capillary rise, when the interface velocities increase, the radial profile progressively flattens compared to the cases at lower interface velocities. 

The progressive reduction of the distance at which the streamwise velocity profile ceases to be parabolic can be correlated with the increase of contact angle for the high Weber number discussed in the previous section. This can be characterized with the degree of parabolicity defined as:
\begin{equation}
    P_{\%}=\frac{|u_{P}-u|}{{u}(0,t)}\times 100\,,
\end{equation}
where $u_{P}$ corresponds to the theoretical parabolic velocity profile. Figure~\ref{fig:parabolicity} shows the evolution of the $L2$-norm of $P_{\%}$, indicated with $P$, normalized with its maximum along the axis of the tube. We define a distance $-\zeta_P$ at which $P/max(P)$ exceeds the standard deviation from the average far from the interface ($\zeta_d<-2$ mm), rounded to $0.1$. The figure shows that after $\zeta_P$ the difference with the parabolic profile increases rapidly, leading to a peak near the interface. The figure also shows the value of $\zeta_P$ gradually reducing with time.
Figure~\ref{fig:parabolicityAll} shows the evolution of $\zeta_P/R$ against $\mathrm{We}$. For increasing $\mathrm{We}$, $\zeta_P/R$ reduces. At higher $\mathrm{We}$, we expect a deformation of the interface shape due to the inertial forces produced by the velocity deceleration. This explains the rapid increase of the contact angle observed in Figure~\ref{fig:HFEContactAngle}.

\section{Conclusions}\label{sec:conclusions} 
This work investigated the capillary-driven flow of HFE7200 and Di-Propylene Glycol in microgravity conditions. The experiments combined backlighting and image processing for interface tracking and PTV for velocimetry. The experiments showed that, despite the different conditions, liquid properties, and wetting behaviour, the shape of the interface remains spherical and can be fully described by its dynamic contact angle. This was shown to be a linear function of the capillary number unless the liquid inertia became dominant. In the case of Di-Propylene Glycol, the flow strongly depends on whether the surface is dry or pre-wet, resulting in different, but constant, interface velocities. This was in line with the hypothesis of an equilibrium between capillary and viscous forces in the liquid column. Conversely, in the case of HFE7200, the capillary rise appears to be driven by the balance of capillary and inertia. In this case, the correlation between contact angle evolution and contact-line velocity is less evident. At the largest contact-line velocities, inertia tends to increase the contact angle and flatten the interface. The PTV measurements show that inertial forces are produced as the liquid at the centre of the channel becomes faster than the meniscus and exerts a dynamic pressure on the interface. 

\section*{References} 
\bibliographystyle{aapmrev4-1}
\bibliography{Main} 

\begin{acknowledgments}
	D. Fiorini is supported by Fonds Wetenschappelijk Onderzoek (FWO), Project number 1S96120N. This work was supported by the ESA Contract No. 4000129315/19/NL/MG. Both partners are gratefully acknowledged.
\end{acknowledgments}

\noindent
\vspace{4mm}
\section*{Data availability}
     The data that support the findings of this study are available on request from the corresponding author. 

\section*{Author declarations}
\noindent
    The authors have no conflicts to disclose.

\end{document}